\documentclass[preprint,showpacs,preprintnumbers,amsmath,amssymb,aps]{revtex4}

\usepackage{graphicx}
\usepackage{dcolumn}
\usepackage{bm}

\begin{document}

\preprint{APS/123-QED}

\title{$S$ = 1/2 ferromagnetic-antiferromagnetic alternating Heisenberg chain in a zinc-verdazyl complex}

\author{Hironori Yamaguchi$^1$, Yasuhiro Shinpuku$^1$,\\ Tokuro Shimokawa$^2$, Kenji Iwase$^1$, Toshio Ono$^1$, Yohei Kono$^3$, \\Shunichiro Kittaka$^3$, Toshiro Sakakibara$^3$, and Yuko Hosokoshi$^1$}
\affiliation{
$^1$Department of Physical Science, Osaka Prefecture University, Osaka 599-8531, Japan \\ 
$^2$Department of Earth and Space Science, Faculty of Science, Osaka University, Osaka 560-0043, Japan\\
$^3$Institute for Solid State Physics, the University of Tokyo, Chiba 277-8581, Japan \\
}

Second institution and/or address\\
This line break forced

\date{\today}

\begin{abstract}
We successfully synthesized the zinc-verdazyl complex [Zn(hfac)$_2$]$\cdot$($o$-Py-V) [hfac = 1,1,1,5,5,5-hexafluoroacetylacetonate; $o$-Py-V = 3-(2-pyridyl)-1,5-diphenylverdazyl], which is an ideal model compound with an $S$ = 1/2 ferromagnetic-antiferromagnetic alternating Heisenberg chain (F-AF AHC). 
$Ab$ $initio$ molecular orbital (MO) calculations indicate that two dominant interactions $J_{\rm{F}}$ and $J_{\rm{AF}}$ form the $S=1/2$ F-AF AHC in this compound. 
The magnetic susceptibility and magnetic specific heat of the compound exhibit thermally activated behavior below approximately 1 K.
Furthermore, its magnetization curve is observed up to the saturation field and directly indicates a zero-field excitation gap of 0.5 T.
These experimental results provide evidence for the existence of a Haldane gap.
We successfully explain the results in terms of the $S=1/2$ F-AF AHC through quantum Monte Carlo calculations with $|J_{\rm{AF}}/J_{\rm{F}}|$ = 0.22. 
The $ab$ $initio$ MO calculations also indicate a weak AF interchain interaction $J'$ and that the coupled
F-AF AHCs form a honeycomb lattice.
The $J'$ dependence of the Haldane gap is calculated, and the actual value of $J'$ is determined to be less than 0.01$|J_{\rm{F}}|$.
\end{abstract}

\maketitle
\section{INTRODUCTION}
Quantum spin systems exhibit unique many-body phenomena caused by strong quantum fluctuations.
According to Haldane's conjecture of 1983~\cite{Haldane}, Heisenberg antiferromagnetic (AF) chain with integer spin values has a finite energy gap between the non-magnetic ground state and the first excited state (Haldane gap), whereas that with half-integer spin values has no energy gap. 
This conjecture has stimulated investigations on Heisenberg AF chains, which established the presence of a Haldane gap for integer spin values~\cite{Todo, Nakano, NENP, MnS2}.
The Haldane state in a Heisenberg AF chain with integer spin values is described as a valence-bond-solid (VBS)~\cite{Affleck, Hagiwara}, where each integer spin is considered as two half-integer spins forming a singlet state between different sites, and the AF correlation function decays exponentially. 
The Haldane state is characterized by the string order parameter defined by $O_{str}^{z} =  \lim_{|i-j| \to \infty}(-{\textless}{S}^{z}_{i} \exp(i{\pi}\sum_{i+1}^{j-1}{S}^{z}_{l}){S}^{z}_{j}{\textgreater})$~\cite{string1, string2}. 
This parameter indicates hidden topological long-range order (LRO) and has a specific value in the Haldane state~\cite{string3, string4}.

The valence-bond picture of the Haldane state can be mapped onto a strong ferromagnetic (F) coupling limit of an F-AF alternating Heisenberg chain (AHC) with half-integer spin values.
Therefore, the $S$ = 1/2 F-AF AHC has been intensively studied in relation with the Haldane state in the $S$ = 1 Heisenberg AF chain.
The thermodynamic properties of the $S$ = 1/2 F-AF AHC were first calculated by Borr$\rm{\acute{a}}$s-Almenar $et$ $al$. nearly two decades ago~\cite{referee}.    
The field-induced phase transitions to a Luttinger liquid (LL) belong to the same universality class as the $S$ = 1 Heisenberg AF chain for both $S$ = 1/2 F-AF and AF-AF AHCs, and the former and the latter have attractive and repulsive interactions between spinless fermions, respectively~\cite{sakai}. 
The dynamical structure factor indicates that the first excited state is located at $q = \pi/2$ for the $S$ = 1/2 F-AF AHC similarly to that for the $S$ = 1 Heisenberg AF chain, while that for $S$ = 1/2 AF-AF AHC is located at $q = 0$~\cite{neut1, neut2}. 
Furthermore, the neutron scattering intensity is predicted to show a peak at an incommensurate position for the $S$ = 1/2 F-AF AHC~\cite{neut1}.
Hida investigated the ground-state properties of the $S$ = 1/2 F-AF AHC for various ratios of exchange constants~\cite{hida1, hida2, hida3}.
There is no discontinuous change in the ground state associated with a phase transition between the Haldane ($|J_{\rm{F}}| \gg J_{\rm{AF}}$) and AF dimer ($|J_{\rm{F}}| \ll J_{\rm{AF}}$) phases.  
The energy gap and string order parameter have finite values for all the ratios of exchange constants and change continuously from the Haldane phase to the AF dimer phase.
It is suggested that the string order parameter represents the strength of the localized singlet correlation, which must be useful to distinguish the VBS from other disordered states.

From the experimental point of view, much effort has been directed at realizing $S$ = 1/2 F-AF AHCs, and several candidates have been reported thus far.
Some of the investigated compounds indicate the disappearance of the Haldane gap and a phase transition to LRO due to relatively strong interchain interactions~\cite{hagiwara, hosokoshi}.  
The copper-based IPA-CuCl$_3$ was considered the ideal example until inelastic neutron scattering (INS) experiments were performed~\cite{IPACu_manaka, IPACu_masuda}.
In fact, the magnetic system in IPA-CuCl$_3$ is characterized as a spin ladder with strongly coupled ferromagnetic rungs~\cite{IPACu_masuda}.
Although an exited energy gap is observed in DMA-CuCl$_3$ through INS experiments, the effective LRO of spins, excluding those forming F-AF AHCs, renders the intrinsic Haldane-like behavior indistinct~\cite{DMA_ajiro,DMA_neutron, DMA_inagaki}.
At present, $\alpha$-CuNb$_2$O$_6$ and Na$_3$Cu$_2$SbO$_6$ are promising candidates with $S$ = 1/2 F-AF AHCs because their magnetic susceptibility, NMR, and INS measurements demonstrate Haldane-like behavior~\cite{CuNbO_1,CuNbO_2, NaCuSbO_1, NaCuSbO_2,NaCuSbO_3}. 
However, the entire phase diagram extending to the gapless phases of these materials have not been experimentally reported, most likely owing to large energy gaps requiring high magnetic fields.   
The entire range of the magnetization curve up to the saturation field, which describes not only the Haldane gap but also the filling of a fermionic band in the field-induced LL phase, is essential to further our understanding of the quantum effect in the crossover from Haldane to AF dimer phases.

In this paper, we successfully synthesized the zinc-verdazyl complex [Zn(hfac)$_2$]$\cdot$($o$-Py-V) [hfac = 1,1,1,5,5,5-hexafluoroacetylacetonate, $o$-Py-V = 3-(2-pyridyl)-1,5-diphenylverdazyl], which is an ideal model compound with the $S$ = 1/2 F-AF AHC, and investigated its crystal structures at room and low temperatures. 
$Ab$ $initio$ molecular orbital (MO) calculations indicated that two dominant interactions form an $S=1/2$ F-AF AHC.
We observed thermally activated behavior associated with an energy gap in the magnetic susceptibility and the magnetic specific heat.
The magnetization curve is observed up to the saturation field and directly indicates a zero-field excitation gap of 0.5 T.
We successfully explained these experimental results in terms of the $S=1/2$ F-AF AHC through quantum Monte Carlo (QMC) calculations. 
Furthermore, we considered the effects of weak AF interchain interactions forming an $S=1/2$ honeycomb lattice.

\section{EXPERIMENTAL}
We synthesized $o$-Py-V through a conventional procedure~\cite{procedure}.
A solution of [Zn(hfac)$_2$]$\cdot$2H$_2$O (176 mg, 0.34 mmol) in 20 ml of heptane was refluxed for 10 min at 70 ${}^\circ$C.
A solution of $o$-Py-V (107 mg, 0.34 mmol) in 10 ml of CH$_2$Cl$_2$ was slowly added, and stirring was continued for 1 h.
After the mixed solution cooled to room temperature, a dark-green crystalline solid was separated by filtration and washed with pentane.
The dark-green residue was recrystallized using CH$_2$Cl$_2$ in an acetonitrile atmosphere (155 mg, 0.195 mmol), after which the yield was approximately 57.4$\%$.

X-ray intensity data were collected using a Rigaku AFC-7R mercury CCD diffractometer and a Rigaku AFC-8R mercury CCD RA-micro7 diffractometer at 293 and 25 K, respectively, with graphite-monochromated Mo K$\rm{\alpha}$ radiation and Japan Thermal Engineering XR-HR10K. 
The structure was solved by a direct method using SIR2004~\cite{SIR2004} and was refined with SHELXL97~\cite{SHELX-97}.
The structural refinement was carried out using anisotropic and isotropic thermal parameters for the nonhydrogen atoms and the hydrogen atoms, respectively. 
All the hydrogen atoms were placed at the calculated ideal positions. 
The magnetic susceptibility and magnetization curves were measured using a commercial SQUID magnetometer (MPMS-XL, Quantum Design) and a capacitive Faraday magnetometer with a dilution refrigerator.
The experimental results were corrected for the diamagnetic contribution of -3.29$\rm{\times}$10$^{-4}$ emu$\cdot$mol$^{-1}$, which was calculated using Pascal's method.
The specific heat was measured with a commercial calorimeter (PPMS, Quantum Design) by using a thermal relaxation method above 2.0 K and by using an adiabatic method between 0.35 K and 2.0 K. 
All experiments were performed using small randomly oriented single crystals with typical dimensions of 2.0$\rm{\times}$1.0$\rm{\times}$0.5 mm$^3$.

$Ab$ $initio$ MO calculations were performed using the UB3LYP method as broken-symmetry (BS) hybrid density functional theory calculations. 
All calculations were performed using the Gaussian 09 program package and 6-31G basis sets.
The convergence criterion was set at 10$^{-8}$ hartree. 
For the estimation of intermolecular magnetic interactions, we applied an evaluation scheme to previously studied multispin systems using the Ising approximation~\cite{MOcal}. 

The QMC code is based on the directed loop algorithm in the stochastic series expansion representation~\cite{QMC2}. 
The calculations for the $S$ = 1/2 F-AF AHC was performed for $N$ = 256 under the periodic boundary condition, where $N$ denotes the system size.
The QMC code for the evaluation of spin gap is based on the multi-cluster loop algorithm in the continuous-time path-integral representation.
The spin gap is calculated by the second-moment method; that is, the gap is the inversed correlation length in the imaginary-time direction.
All calculations were carried out using the ALPS application~\cite{ALPS, Todo, ALPS3}.

\section{RESULTS}

\subsection{Crystal structure and magnetic model}
The crystallographic data are summarized in Table I~\cite{crystal}, and the molecular structure is shown in Fig. 1(a).
The verdazyl ring (which includes four nitrogen atoms), the two upper phenyl rings, and the bottom pyridine ring are labeled as ${\rm{R}_{1}}$, ${\rm{R}_{2}}$, ${\rm{R}_{3}}$, and ${\rm{R}_{4}}$, respectively.
The MO calculation indicates that approximately 58\% of the total spin density is present on ${\rm{R}_{1}}$. 
While ${\rm{R}_{2}}$ and ${\rm{R}_{3}}$ each account for approximately 17 \% of the relatively large total spin density, ${\rm{R}_{4}}$ accounts for less than 6 \% of the total spin density.
Since Zn(hfac)$_2$ has a low spin density, it works as a spacer between verdazyl radicals, resulting in the low dimensionality of the magnetic lattice. 
We focus on the structural features related to the $o$-Py-V to consider intermolecular interactions.
We evaluated the intermolecular magnetic interactions of all molecular pairs within 4.0 $\rm{\AA}$ at both 293 and 25 K through the $ab$ $initio$ MO calculations.
Consequently, we found that there are three types of dominate interactions related to M$_0$-M$_1$, M$_0$-M$_2$, and M$_0$-M$_3$ molecular pairs, as shown in Figs. 1(b)-(d).
The M$_0$-M$_1$ molecular pair has an N-C short contact $d_1$, which is doubled by inversion symmetry, as shown in Fig. 1(b), and it is approximately 3.48 $\rm{\AA}$ at both 293 and 25 K. 
A strong F interaction $J_{\rm{F}}$ is evaluated between these molecules. 
The M$_0$-M$_2$ molecular pair has a C-C short contact $d_2$, which is also doubled by inversion symmetry, as shown in Fig. 1(c).
The values of $d_2$ at 293 and 25 K are 3.48 and 3.39 $\rm{\AA}$, respectively.
A relatively strong AF interaction $J_{\rm{AF}}$ is evaluated between these molecules.
The M$_0$-M$_1$ and M$_0$-M$_2$ molecular pairs are alternately aligned along the $c$-axis, as shown in Fig. 1(e), and form an $S$ = 1/2 F-AF AHC consisting of the $J_{\rm{F}}$ and the $J_{\rm{AF}}$, as shown in Fig. 1(f). 
The M$_0$-M$_3$ molecular pair, which is related by inversion symmetry, is associated with interchain interaction $J'$ and has a C-C short contact $d_3$.
The values of $d_3$ at 293 and 25 K are 3.51 and 3.44 $\rm{\AA}$, respectively.
Although there is relatively short contact, the small overlap of the $\pi$ orbitals, which expand perpendicular to the planes, should render $J'$ weak. 
This interchain interaction forms a honeycomb lattice in the $ac$-plane.
The evaluated values of the exchange interactions are summarized in Table II.
They are defined with the Heisenberg spin Hamiltonian:
\begin{equation}
\mathcal {H} = J_{\rm{F}}{\sum^{}_{i}}\textbf{{\textit S}}_{2i}{\cdot}\textbf{{\textit S}}_{2i+1}+J_{\rm{AF}}{\sum^{}_{i}}\textbf{{\textit S}}_{2i-1}{\cdot}\textbf{{\textit S}}_{2i}+J'{\sum^{}_{kl}}\textbf{{\textit S}}_{k}{\cdot}\textbf{{\textit S}}_{l},
\end{equation}
where $\textbf{{\textit S}}$ is an $S$=1/2 spin operator. 
It is remarkable that the absolute value of the interchain interaction $J'$ at 25 K has a small value, while the other interactions are more dominant at 25 K than at 293 K. 
Thus, the one-dimensionality is expected to be enhanced in the low-temperature regions, which should provide an intrinsic magnetic behavior reflecting the ground state of the $S$ = 1/2 F-AF AHC.

\begin{table}
\caption{Crystallographic data for [Zn(hfac)$_2$]$\cdot$($o$-Py-V)}
\label{t1}
\begin{center}
\begin{tabular}{ccc}
\hline
\hline 
Formula & \multicolumn{2}{c}{C$_{29}$H$_{18}$F$_{12}$N$_{5}$O$_{4}$Zn}\\
Crystal system & \multicolumn{2}{c}{Monoclinic}\\
Space group & \multicolumn{2}{c}{$P$2$_1$/$c$} \\
Temperature (K) & 293(2) & 25(2)\\
Wavelength ($\rm{\AA}$) & \multicolumn{2}{c}{0.7107} \\
$a (\rm{\AA}$) &  9.130(5) &  8.8144(17) \\
$b (\rm{\AA}$) &  31.839(16) & 31.582(6) \\
$c (\rm{\AA}$) & 10.885(6) & 10.826(2)\\
$\beta$ (degrees) &  93.286(5) & 92.772(4)\\
$V$ ($\rm{\AA}^3$) & 3159(3) & 3010.3(10) \\
$Z$ & \multicolumn{2}{c}{4} \\
$D_{\rm{calc}}$ (g cm$^{-3}$) & 1.669 & 1.752\\
Total reflections & 4796 & 4390\\
Reflection used & 4289 & 3991\\
Parameters refined & \multicolumn{2}{c}{460}\\
$R$ [$I>2\sigma(I)$] & 0.0661 & 0.0787\\
$R_w$ [$I>2\sigma(I)$] & 0.1669 & 0.1922\\
Goodness of fit & 1.065 & 1.029\\
CCDC & 1020097 & 1020098 \\
\hline
\hline
\end{tabular}
\end{center}
\end{table}

\begin{table}
\caption{Evaluated magnetic interactions through the $ab$ $initio$ MO calculations at 293 and 25 K and the experimental analysis by using QMC calculations}
\label{t1}
\begin{center}
\begin{tabular}{c c c c c c}
\hline 
Method &   $|J_{\rm{AF}}/J_{\rm{F}}|$   & $|J'/J_{\rm{F}}|$ & $J_{\rm{F}}/k_{\rm{B}}$ (K) &  $J_{\rm{AF}}/k_{\rm{B}}$ (K) &  $J'/k_{\rm{B}}$ (K) \\
\hline 
MO at 293 K  & 0.29 & 0.13 & -17.2 & 5.0 & 2.3 \\
MO at 25 K & 0.34 & 0.05 & -21.2 & 7.2 & 1.1 \\
QMC analysis & 0.22 & $\textless$ 0.01 & -12.8 & 2.8 & $\textless$ 0.1 \\
\hline
\end{tabular}
\end{center}
\end{table}

\begin{figure}[t]
\begin{center}
\includegraphics[width=20pc]{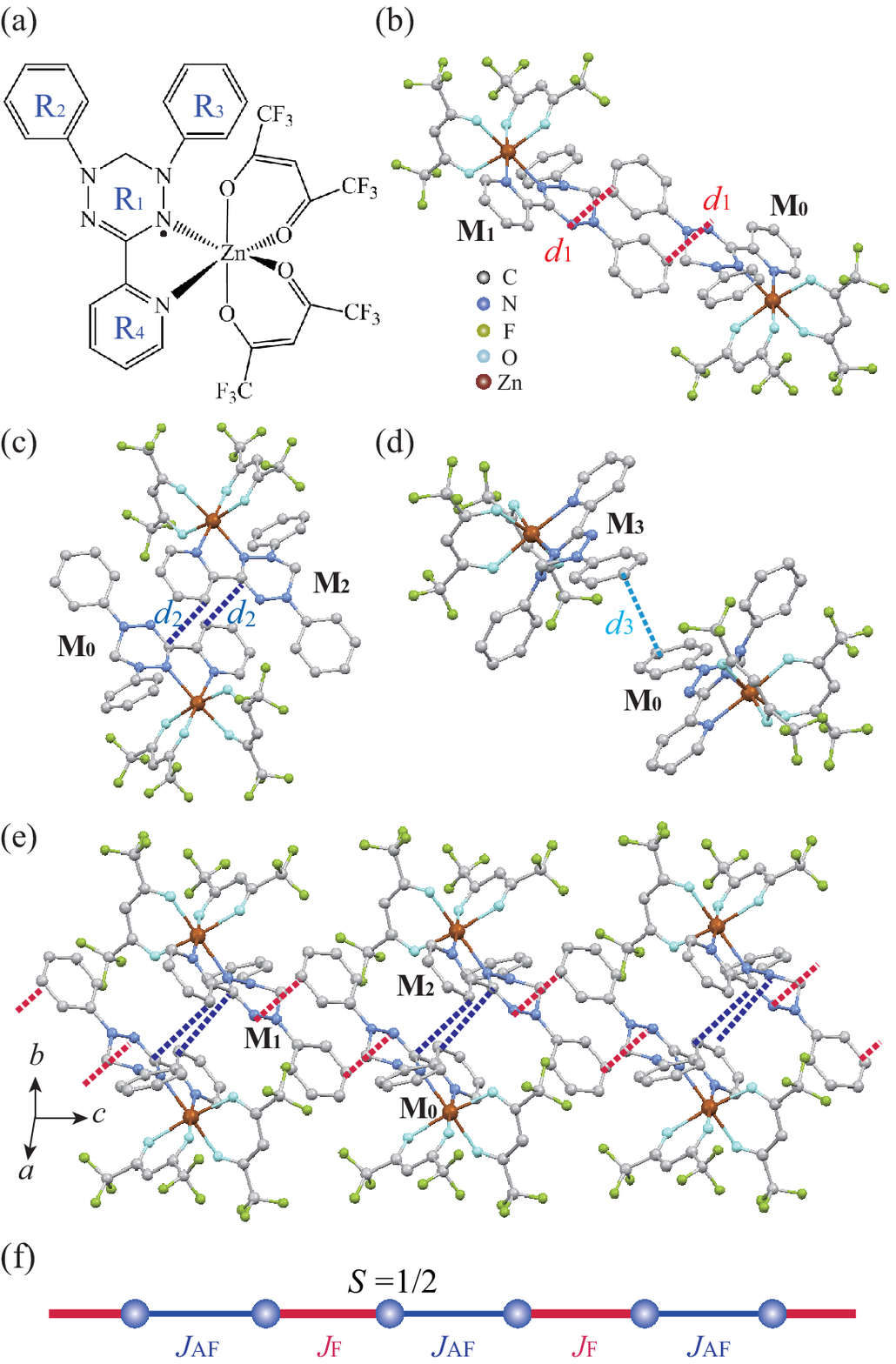}
\caption{(color online) (a) Molecular structure of [Zn(hfac)$_2$]$\cdot$($o$-Py-V). Molecular packing of neighboring (b) M$_0$-M$_1$, (c) M$_0$-M$_2$, and (d) M$_0$-M$_3$ molecular pairs, which are associated with $J_{\rm{F}}$, $J_{\rm{AF}}$, and $J'$, respectively. Broken lines indicate C-C and N-C short contacts. Hydrogen atoms are omitted for clarity. (e) Crystal structure forming an alternating chain consisting of $J_{\rm{F}}$ and $J_{\rm{AF}}$ along the $c$-axis, and (f) the corresponding $S$ = 1/2 F-AF AHC. }\label{f1}
\end{center}
\end{figure}

\subsection{Magnetic and thermodynamic properties}
Figure 2 shows the temperature dependence of the magnetic susceptibility ($\chi$ = $M/H$) at 0.1 T ($H$ = 1000 Oe).
We observed a broad peak at approximately 1.0 K, below which $\chi$ decreases with decreasing temperatures, indicating the existence of a nonmagnetic ground state separated from the excited states by an energy gap.
Above 100 K, it follows the Curie-Weiss law, and the Weiss temperature is estimated to be ${\theta}_{\rm{W}}$ = +3.0(3) K, which indicates the dominant contribution of $J_{\rm{F}}$. 
We evaluated the paramagnetic impurities to be approximately 1.2\% of all spins, which is defined to fit the following calculated result and is close to those evaluated in other verdazyl radical crystals~\cite{3Cl4FV, 3Br4FV, 26Cl2V}, and subtracted it from the raw data by assuming conventional paramagnetic behavior, as shown in Fig. 2.
The contribution of $J_{\rm{F}}$ appears in the temperature dependence of $\chi$$T$, which increases with
decreasing temperatures down to approximately 12 K, as shown in the inset of Fig. 2. 
Below 12 K, $\chi$$T$ decreases with decreasing temperature, indicating the existence of AF interactions.

Figure 3 shows the magnetization curve at 0.13 K.
We subtracted a small paramagnetic contribution given by the Brillouin function in the low-field region that is evaluated to account for 1.2$\%$ of all spins.
The saturation value of 0.98 $\mu_{\rm{B}}$/f.u. indicates that the purity of the radicals is approximately 98 $\%$, and we consider this purity in the following analysis. 
The corrected magnetization curve without the paramagnetic contribution clearly exhibits a small zero-field excitation gap of approximately 0.5 T, as shown in Fig. 3 and its upper inset.
The lower inset of Fig. 3 shows the field derivative of the corrected magnetization curve ($dM/dB$).
The steep increase of $dM/dB$ at the critical field region evidences the gapped behavior. 
This gapped behavior of the magnetization curve strongly suggests the realization of a Haldane state in the $S$ = 1/2 F-AF AHC.

The inset of Fig. 4 shows the experimental result of the total specific heat $C_{\rm{p}}$ at 0 and 1.0 T.
The magnetic specific heats $C_{\rm{m}}$ are obtained by subtracting the lattice contribution by assuming Debye's $T^3$-law as $0.020T^3$ (J/mol K), which corresponds to about 23 K of the Debye temperature. 
There is no sharp peak associated with a phase transition to an ordered state.
A clear Schottky-like peak associated with an excitation gap below approximately 1.0 K at zero field is observed, which is consistent with the existence of a Haldane gap.

\begin{figure}[t]
\begin{center}
\includegraphics[width=20pc]{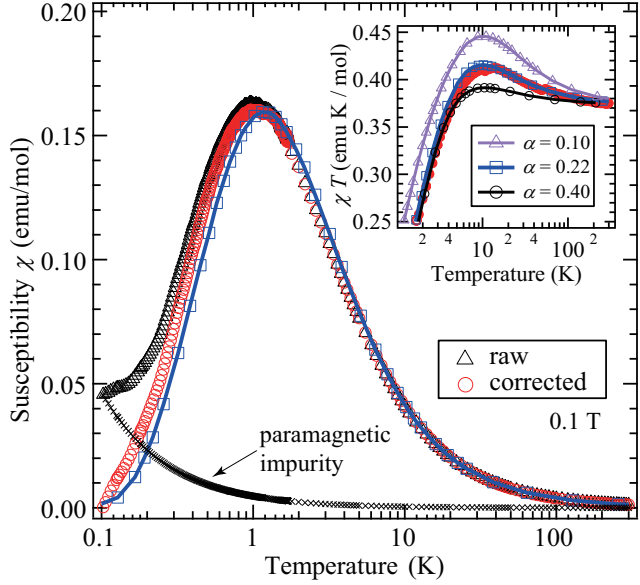}
\caption{(color online) Temperature dependence of magnetic susceptibility ($\chi$ = $M/H$) of [Zn(hfac)$_2$]$\cdot$($o$-Py-V) at 0.1 T. The open circles denote raw data, and the open triangles are corrected for the paramagnetic term due to the impurity. The inset shows the temperature dependence of ${\chi}T$. The solid lines with open triangles, squares, and circles represent the calculated results for the $S$ = 1/2 F-AF AHC with $\alpha = |J_{\rm{AF}}/J_{\rm{F}}|$ = 0.10, 0.22, and 0.40, respectively. }\label{f2}
\end{center}
\end{figure}

\begin{figure}[t]
\begin{center}
\includegraphics[width=20pc]{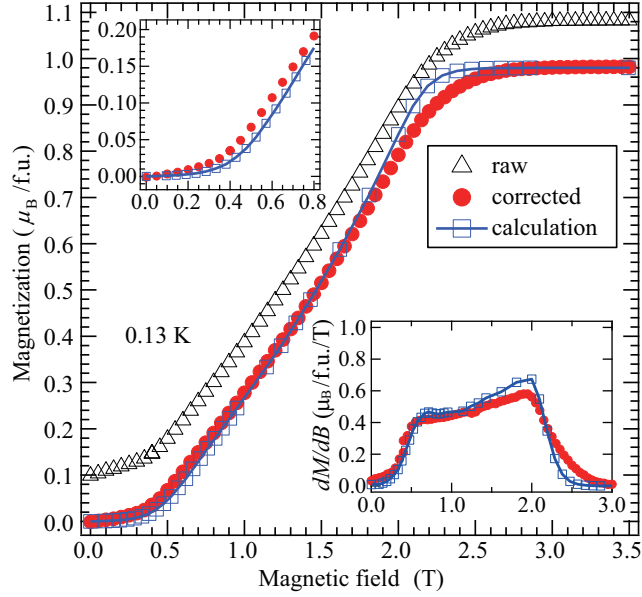}
\caption{(color online) Magnetization curve of [Zn(hfac)$_2$]$\cdot$($o$-Py-V) at 0.13 K. The open triangles denote raw data, and the values of the vertical axis have been shifted by 0.1 $\mu_B$/f.u. for clarity. The closed circles are corrected for the paramagnetic term due to the impurity. The solid line with open squares represents the calculated result for the $S$ = 1/2 F-AF AHC with $\alpha = |J_{\rm{AF}}/J_{\rm{F}}|$ = 0.22. The upper and the lower insets show the expansion of the low-field region and field derivative $dM/dB$, respectively. }\label{f3}
\end{center}
\end{figure}

\begin{figure}[t]
\begin{center}
\includegraphics[width=20pc]{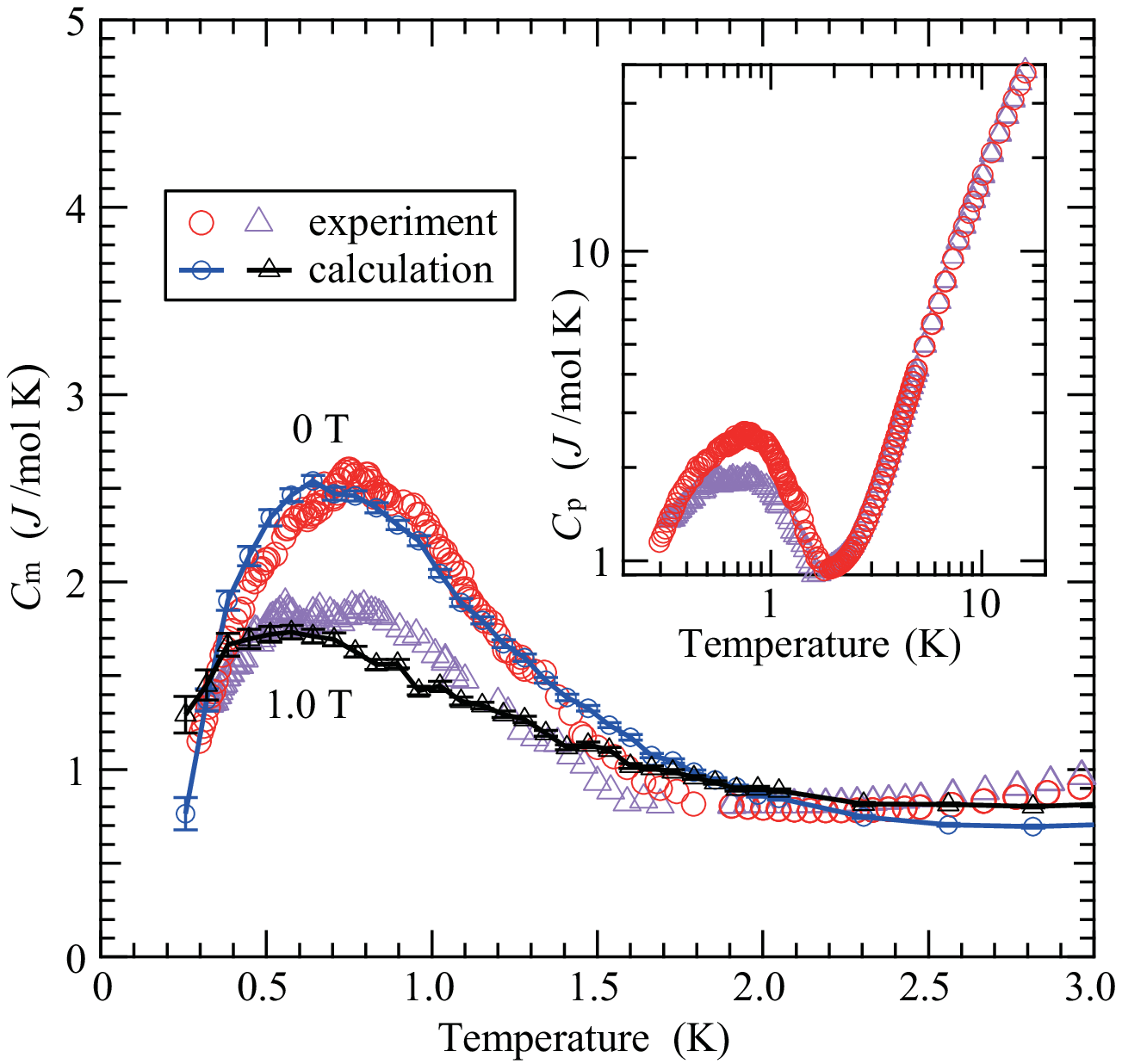}
\caption{(color online) Temperature dependence of magnetic specific heat of [Zn(hfac)$_2$]$\cdot$($o$-Py-V) at 0 and 1.0 T. The solid lines with open circles and triangles represent the calculated results for the $S$ = 1/2 F-AF AHC with $\alpha = |J_{\rm{AF}}/J_{\rm{F}}|$ = 0.22 at 0 and 1.0 T, respectively. The inset shows the total specific heat.}\label{f4}
\end{center}
\end{figure}

\section{DISCUSSION}
We discuss the ground state of the present spin model.
The MO calculations show that the dominant intermolecular interactions $J_{\rm{F}}$ and $J_{\rm{AF}}$ form the $S$ = 1/2 F-AF AHC, and the one-dimensionality is enhanced in the low-temperature regions, as summarized in Table II. 
Accordingly, we analyzed the experimental results in terms of the $S$ = 1/2 F-AF AHC as shown in Fig. 1(f). 
We calculated the magnetic susceptibility, magnetization curve, and magnetic specific heat as a function of $\alpha=|J_{\rm{AF}}/J_{\rm{F}}|$ by using the QMC method.
As is often the case with conventional radical compounds, we assume the Heisenberg spin Hamiltonian and $g$ = 2.00.
Although there is no distinct dependence of ${\chi}$ on $\alpha$, the maximum value of ${\chi}$ $T$ is very sensitive to the change in $\alpha$, as shown in the inset of Fig. 2. 
We obtained good agreement between the experiment and calculation by using the parameters $J_{\rm{F}}/k_{\rm{B}}$ = -12.8 K and $J_{\rm{AF}}/k_{\rm{B}}$ = 2.8 K ($\alpha$ = 0.22), as shown in Fig. 2 and its inset.
The slight difference, especially in the low-temperature region, might originate from the inaccuracy in the subtraction of the paramagnetic impurity.
In addition, the calculated result corresponds to the limiting behavior $H \to 0$, which also causes a slight difference reflecting the change in the energy gap at the experimental field of 0.1 T. 
A comparison between the experimental results and $ab$ $initio$ calculations on verdazyl-based materials~\cite{fine-tune,26Cl2V} shows that the obtained parameters are consistent with those evaluated from the MO calculation.
The magnetization curves calculated using the QMC method also well reproduced the experimental results using the same parameters, as shown Fig. 3 and its insets.
Because the energy scale of the exchange interactions is relatively small, the finite temperature effect is enhanced in the present system, resulting in the finite slope in the gapped region below 0.5 T and the gradual change at the critical field.
We find a slight deviation in the vicinity of the saturation field. 
In the experimental result, the nonlinear behavior originating from the one-dimensional (1D) quantum fluctuation is marginally suppressed by the weak interchain interactions forming the two-dimensional (2D) honeycomb lattice, resulting in the difference between the experiment and calculation.
The calculated result of the magnetic specific heat also well reproduces the Schottky-like peak at 0 T, as shown in Fig. 4.
Considering the gapless phase above 0.5 T, the broad peak observed at 1.0 T indicates the development of 1D short-range order.  
These different origins of the broad peaks indeed give rise to a large difference in the peak shapes and values.
The difference between the experiment and calculation should arise from the inaccuracy in the subtraction of the lattice contribution, which we assumed largely as Debye's $T^{3}$-law.
The ground state for the $S$ = 1/2 F-AF AHC with $\alpha$ = 0.22 is a Haldane state with the excitation gap $\Delta$ $\approx$ 0.67 K to the lowest triplet states, which is consistent with a previous theoretical study~\cite{hida1}. 
The corresponding string order parameter is predicted to show an intermediate value between the $S$ = 1 Haldane and $S$ = 1/2 AF dimer phases~\cite{hida1}.
The gapless phase is described as an LL and expected to have attractive spinless fermions~\cite{sakai}.   
The present small excitation gap and easily accessible saturation field will enable quantitative tests of such properties through various measurements.

Finally, we consider the contribution of $J'$ quantitatively.
The slight deviations in the low-temperature regions of the magnetic susceptibility and in the vicinity of the zero and
the saturation fields of the magnetization curve may indicate slight contributions from the weak interchain interaction $J'$.
The $S$ = 1/2 F-AF AHCs coupled by the AF $J'$ form an $S$ = 1/2 honeycomb lattice, as shown in Fig. 5.
We calculated the value of the energy gap $\Delta$ as a function of $\beta$ = $|J'/J_{\rm{F}}|$, assuming $\alpha$ = 0.22 and $J_{\rm{F}}/k_{\rm{B}}$ = -12.8 K using the QMC method at $|T/J_{\rm{F}}|$ = 0.001.
The inset of Fig. 5 shows the system size dependence of $\Delta$ for each value of $\beta$.
Assuming an energy gap in a 2D system, finite-size data are fitted using $\Delta_{N} = \Delta_{\infty} + C {\rm{exp}}(-D N^{1/2})$, where $C$ and $D$ are fitting parameters~\cite{gap1, gap2}. 
It is clear that the fitting curves are appropriate for $\beta$ $\le$ 0.02, as shown in the inset of Fig. 5. 
For $\beta$ $\ge$ 0.03, on the other hand, it is difficult to estimate the value of the energy gap in the thermodynamic limit by using the above fitting equation because there is no convex downward behavior in the finite-size data up to $N$=576.
Therefore, we roughly evaluated the energy gap in the thermodynamic limit using a linear fit with the values of the four largest system sizes ($N=$324, 400, 484, and 576) for $\beta$$\ge$ 0.03.
The energy gap decreases dramatically with increasing $\beta$ and almost disappears at $\beta$ = 0.04, resulting in the phase transition to an AF ordered state, as shown in Fig. 5.
The experimental result of the magnetization curve shows that $\Delta$ $\approx$ 0.5 T ($\approx$ 0.67 K); the experimental and calculated results are consistent.
Therefore, even if the interchain interaction is considered, the value of $\Delta$ should be more than 0.4 T ($\approx$ 0.54 K), as deduced from the gapped behavior of the magnetization curve.
Considering that the calculated value of $\Delta$ for $\beta$ = 0.01 is approximately 0.51 K, the actual value of $\beta$ is roughly estimated to be less than 0.01.

\begin{figure}[t]
\begin{center}
\includegraphics[width=20pc]{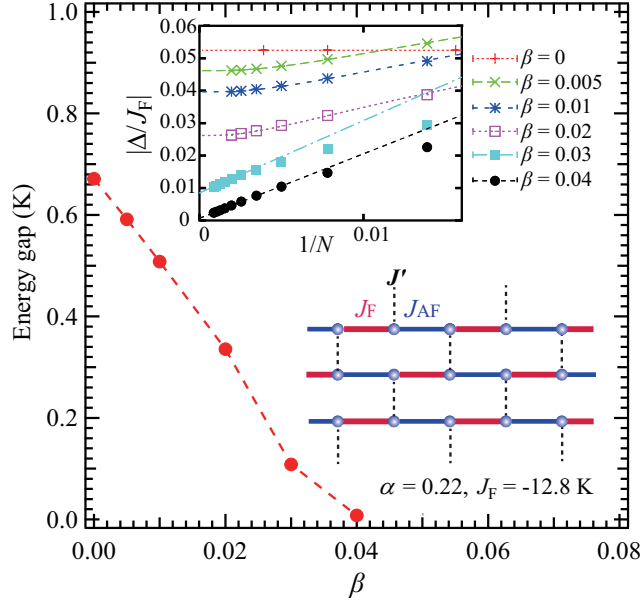}
\caption{(color online) $\beta$ = $|J'/J_{\rm{F}}|$ dependence of the energy gap $\Delta$ for the weakly coupled $S$ = 1/2 F-AF AHCs with $\alpha = |J_{\rm{AF}}/J_{\rm{F}}|$ = 0.22 and $J_{\rm{F}}/k_{\rm{B}}$ = -12.8 K. The inset shows the 1/$N$ dependence of $|\Delta/J_{\rm{F}}|$ for each $\beta$. The broken lines are fitting curves described in the text. The illustration shows the $S$ = 1/2 honeycomb lattice consisting of $J_{\rm{F}}$, $J_{\rm{AF}}$, and $J'$.}\label{f3}
\end{center}
\end{figure}

\section{SUMMARY}
We succeeded in synthesizing a zinc-verdazyl complex [Zn(hfac)$_2$]$\cdot$($o$-Py-V).
The $ab$ $initio$ MO calculations indicate the existence of dominant interactions $J_{\rm{F}}$ and $J_{\rm{AF}}$ forming an $S=1/2$ F-AF AHC in this compound.
Thermally activated behaviors are observed in the magnetic susceptibility and the magnetic specific heat, and the magnetization curve indicates the existence of an energy gap of 0.67 K.
These features are described as the magnetic behavior in an $S=1/2$ F-AF AHC with the parameters $J_{\rm{F}}/k_{\rm{B}}$ = -12.8 K and $J_{\rm{AF}}/k_{\rm{B}}$ = 2.8 K ($|J_{\rm{AF}}/J_{\rm{F}}|$ = 0.22) by using the QMC method. 
Furthermore, the MO calculation also indicated the existence of the interchain interaction $J'$ and that the coupled F-AF AHCs form a honeycomb lattice. 
We consider the contribution of $J'$ to the value of the Haldane gap by using the QMC method, and the actual value of $|J'/J_{\rm{F}}|$ is evaluated to be less than 0.01. 
These results demonstrate that [Zn(hfac)$_2$]$\cdot$($o$-Py-V) is an ideal model compound with an $S=1/2$ F-AF AHC.
Further investigations of dynamical properties will yield quantitative information about the intermediate phase between $S$ = 1 Haldane and $S$ = 1/2 AF dimer phases and the field-induced LL phase of attractive spinless fermions.

\begin{acknowledgments}
We thank T. Tonegawa, S. Yasuda, and S. Todo for the valuable discussions. This research was partly supported by KAKENHI (Nos. 24740241, 24540347, and 24340075), the CASIO Science Promotion Foundation, and Grant for Basic Science Research Projects from the Sumitomo Foundation．
A part of this work was performed as the joint-research program of ISSP, the University of Tokyo and the Institute for Molecular Science. 
Some computations were performed using the facilities of the Supercomputer Center, ISSP, the University of Tokyo. 
\end{acknowledgments}


\end{document}